\documentclass[runningheads]{llncs}
\usepackage[T1]{fontenc}
\usepackage{graphicx}
\usepackage{subcaption} 
\usepackage[export]{adjustbox}
\usepackage{tabularx}
\usepackage{tabularx,makecell,array}
\usepackage[dvipsnames]{xcolor}
\definecolor{red}{rgb}{1.0, 0.3, 0.3}
\definecolor{gold}{rgb}{0.8, 0.6, 0.2}
\definecolor{blue}{rgb}{0.25, 0.5, 1}
\definecolor{brown}{rgb}{0.6, 0.4, 0.2}
\usepackage{cite}

\begin{document}
\title{Closing the Gap Between Float and Posit Hardware Efficiency}
%
%
\author{Aditya Anirudh Jonnalagadda\inst{1}\orcidID{0000-0003-3011-1273} \and
Rishi Thotli\inst{1} \and
John~L.~Gustafson\inst{2}\orcidID{0000-0002-2957-1304}}
\authorrunning{A. Jonnalagadda et al.}
%
\institute{Birla Institute of Technology and Science, Pilani, Hyderabad Campus, Hyderabad, Telangana 500078, India
\email{adi.anirudh1610@gmail.com}\
\email{rishithotli42@gmail.com}\\
\and
Arizona State University, Tempe, Arizona USA\\
\email{\{jlgusta6\}@asu.edu}}
\maketitle              
\begin{abstract}
  The \textit{b-posit}, or bounded posit, is a variation of the posit format designed for high performance computing (HPC) and AI applications. Unlike traditional floating-point formats (floats, for short), posits use variable-length fields for exponent scaling and significand, providing a better match to application requirements for the same bit width. However, this flexibility introduces high worst-case overhead in the decode-encode logic when all bits express the exponent scaling, an overhead that exceeds the cost of handling subnormals for floats. To address this, the b-posit restricts the regime field to a 6-bit limit, which reduces the variability in regime and fraction sizes. If we choose an exponent size \textit{eS} of 5 bits, the dynamic range is $2^{-192}$ to $2^{192}$ (about $10^{-58}$ to $10^{58}$) and the quire size is 800 bits, for any precision $n>12$. This constraint improves the numerical properties of the format and simplifies hardware implementation by allowing decode-encode operations to be handled with basic multiplexers, as a result of the limited number of possible regime sizes. Our newly designed 32-bit b-posit decoder circuits achieve significant improvements (79\% less power consumption, 71\% smaller area, and a 60\% reduction in latency) compared to standard posit decoders. Likewise, the 32-bit b-posit encoder shows 68\% lower power usage, 46\% less area, and 44\% shorter delay. The proposed b-posit hardware exhibits superior scalability with increasing bit widths, outperforming standard posit hardware at higher precisions, including 64-bit formats where the advantages are even greater. Notably, the proposed b-posit decode-encode hardware matches or exceeds the performance of IEEE compliant 32-bit floating-point, offering a faster and smaller area implementation, with a slight increase in worst-case power consumption due to the higher speed. The b-posit hardware design allows the clean mathematical behavior and higher accuracy of posits versus (16-, 32-, and 64-bit) IEEE floats without the costs of power, area, or latency that have been observed for posits that follow the Posit™ Standard (2022). We believe that the b-posit should influence future revisions of that standard.
      \keywords{Posit format  \and Computer arithmetic \and Floating-point format}.
\end{abstract}
\section{The Evolution of Posit Exponent Size Choices}
\subsection{Overview of Standard Float and Posit Formats}

The posit number format [1, 2, 3], is a Type III unum proposed as an alternate data type to the IEEE 754 floating-point standard for computing with real numbers [4]. The posit format superficially resembles the floating-point format, but the biased exponent is replaced with a regime and (unbiased) exponent combination where the regime can vary in length and is interpreted as a sort of signed unary;  Fig. 1 compares 16-bit IEEE floats with 16-bit standard posits. While standard floating-point numbers (\textit{floats} here, for short) map to the sign-magnitude integer representation that computers used in the 1950s, posits map to modern signed integers with 2's complement representation.
\begin{figure}[ht]
    \centering
    \includegraphics[width=0.9\linewidth]{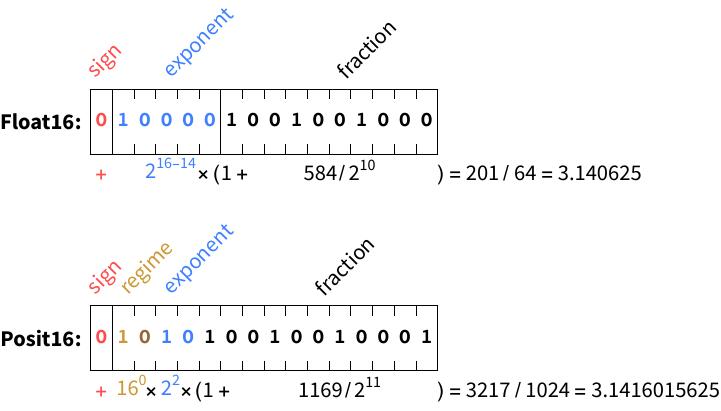}
    \caption{Comparison of 16-bit IEEE float vs standard posit for representing $\pi$}
    \label{fig:floatsvsposits}
\end{figure}

To make bit strings easier to read, the bits for each posit field are color coded \textbf{\textcolor{red}{red}}, \textbf{\textcolor{gold}{gold}}, \textbf{\textcolor{blue}{blue}}, and \textbf{black} respectively, following the convention of [1, 2, 3]. The opposite bit terminating the regime is color coded \textbf{\textcolor{brown}{brown}}. The posit approximation for $\pi$ is over $100$ times more accurate than the float approximation.

The most significant bit (MSB) of a posit indicates its sign, \texttt{1} for negative values and \texttt{0} for zero and positive values. While this looks similar to floats, the fact that floats use sign-magnitude arithmetic instead of 2's complement means that negative float values are in the reverse order of negative posit values.

The regime field consists of a run of $k$ identical bits terminated by the first opposite bit. If the run is of $k$ \texttt{0} bits, the regime value $r$ is $-k$. If the run is $k$ \texttt{1} bits, the regime value $r$ is $k-1$. See Fig.~2, where the ``$\cdot$'' bits are not part of the regime and can be used for other purposes.

\begin{figure}[ht]
    \centering
    \includegraphics[width=0.4\linewidth]{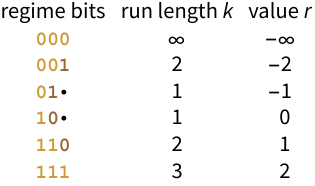}
    \caption{Example values with a maximum of three regime bits}
    \label{fig:3bitregime}
\end{figure}

Beyond the $n$ expressed bits, there is always an implied infinite sequence of \texttt{0} bits, just as a decimal like $3.75$ implies $3.75000\cdots$. The implied bits are \textit{ghost bits}. If the regime is a run of \texttt{0}-valued bits that extends to the least significant bit (LSB) of the number, then the run is infinitely long and therefore represents $-\infty$. If the run of bits is \texttt{1} bits all the way to the LSB, the next bit is a ghost bit and hence has value \texttt{0}, thus serving as the opposite bit. Ghost bits can be regime, exponent, or fraction bits (see Fig.~3).

\begin{figure}[ht]
    \centering
    \includegraphics[width=0.8\linewidth]{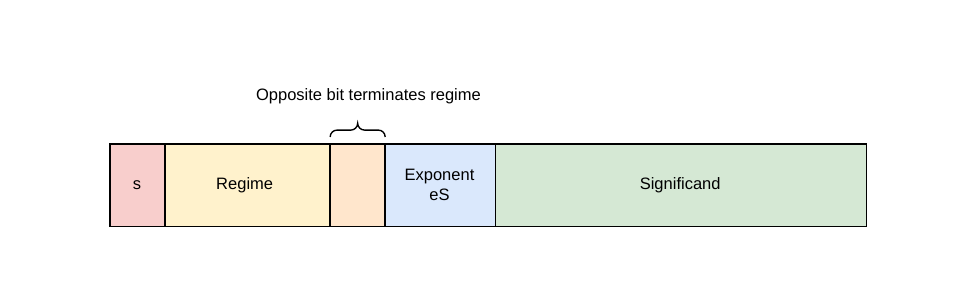}
    \caption{Bits inside the box are explicit bits; trailing bits are ghost bits}
    \label{fig:ghostbits}
\end{figure}

The regime format resembles unary (also called thermometer code) notation, but with 2's complement to indicate both negative and positive integers \textit{within a fixed number of bits}. It has been compared to Rice-Golomb encoding, but Rice-Golomb is for unsigned integers and assumes no limit on the number of expressed bits.

Fig.~4 shows the fields of a generic posit format. The sign, regime and exponent fields determine the effective exponent $T$, which is $(r \cdot 2^{eS} + e)$ for positive values ($s = 0$) and $-(r \cdot 2^{eS} + e + 1)$ for negative values ($s = 1$). Therefore, a unified expression for $T$ is
\[
T = (1 - 2s) \cdot (r \cdot 2^{eS} + e + s).
\]

The remaining bits of the $n$-bit word constitute the fraction $f$, $0\leq~f<1$. For positive posits, the significand is $1 + f$, so it is in the half-open interval $[1, 2)$. For negative posits, the significand is $-2 + f$ and thus is in the half-open interval $[-2, 1)$. This allows posits to perfectly map to (2’s complement) signed integers. In other words, the significand is $1 - 3s + f$, where $s$ is the value of the sign bit.

\begin{figure}[ht]
    \centering
    \includegraphics[width=1\linewidth]{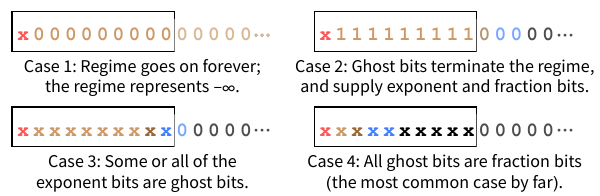}
    \caption{Representation of a General Posit}
    \label{fig:generalposit}
\end{figure}

If the bits are all zero after the sign bit, then if $s = 0$ the above formula is 
\[
X = 2^{-\infty} = 0;
\] 
if $s = 1$, 
\[
X = -2^{\infty} = -\infty,
\] 
which is not a real number. That bit pattern is ``NaR'' for ``Not a Real'' and is used for all exceptions such as infinities, $0/0$, and complex values such as $\sqrt{-1}$. In comparison operations, NaR is less than any other posit, so the comparison hardware for signed integers suffices to compare posits. This alone represents a significant savings in chip area compared to the floating-point format.

Unlike floats, posits employ variable-length fields, where the sizes of the effective exponent and significand change dynamically based on the value being represented. This enables tapered relative accuracy, where numbers near unity benefit from additional significand bits borrowed from the regime, while representations of very large magnitudes allocate more bits to the regime at the expense of significand precision. Although tapered precision is useful for increasing dynamic range and improving the relative accuracy of the most commonly used real values, variable length encoding introduces hardware complexity, since each field must be decoded dynamically to form a float-like internal representation. Once the decoding is complete and the arithmetic operation is performed, the results are to be encoded or packed back into the posit format. This hardware cost of standard posit decode-encode is greater than that of exception detection and subnormal handling of floats; as a result, a posit arithmetic block comes at a greater cost than its floating-point analogue, as shown in \cite{ref6,ref7,ref10}.

\subsection{The Initial Guess for Exponent Size }
The choice of the number of bits in the exponent field determines the dynamic range. Each increment of the exponent field size squares the dynamic range, at the expense of one less bit of accuracy in the most accurate significands. A too small dynamic range is catastrophic in general, and a too large dynamic range will have reduced accuracy and some bit patterns that are rarely actually used. This choice was the biggest source of uncertainty in the original 2017 paper that introduced the posit format.

Like the Posit standard, here we use \texttt{ lowerCamelCase} for variable names with more than one letter, so \texttt{ exponentSize} (number of bits) is abbreviated as $eS$; it can be any nonnegative integer $0, 1, 2, \cdots$. The original paper (2017) made a strawman guess about the appropriate size of $eS$ as a function of $n$, shown in Table~3.

\begin{table}[h]
\centering
\begin{tabular}{|c|c|}
\hline
\textbf{Posit Precision $n$} & \textbf{Exponent Size \textit{eS}} \\ \hline
8   & 0       \\ \hline
16  & 1       \\ \hline
32  & 2       \\ \hline
$2^n$ & $n-3$ \\ \hline
\end{tabular}
\caption{The 2017 strawman proposal for exponent sizes}
\label{tab:es-values}
\end{table}

The guess was based on the hunch that having a longer significand is more important than a large dynamic range. Software simulators such as Cerlane Leong's SoftPosit and Theo Omtzigt's Universal were constructed to explore which choices gave the most accurate results for actual workloads.

\subsection{Refinement: Constant Exponent Size for All Precisions} 

Years of experiments on a wide range of applications (AI, signal processing, HPC) showed that the best results were usually for $\textit{eS} = 2$ or $\textit{eS} = 3$, independent of the precision $n$. Fixing the value of $\textit{eS}$ solved two problems:

\begin{itemize}
    \item Changing the precisions became trivial, without the need to decode the fields first.
    \item $n$ is not limited to integer powers of 2 but can be any integer 2 or greater.
\end{itemize}

Hence, after five years of empirical studies, the Posit Working Group ratified the \textit{Posit™ Standard (2022)} specifying $eS = 2$. The value of the exponent is $e$, treating its bits as an unsigned integer.

After several more years of working with standard posits, it is clear that there are two drawbacks to this choice. First, as precision $n$ grows, the dynamic range grows without limit, and for $64$-bit posits the magnitudes range from $2^{-248}$ to $2^{248}$ (about $10^{-75}$ to $10^{75}$), larger than needed even for cosmological applications.

Second, evaluating the worst-case run length in the regime is expensive. The latency to count leading \texttt{0} bits or leading \texttt{1} bits grows as the logarithm of the number of bits with optimal circuits;  some naive designs use linear cost methods, but divide-and-conquer methods for run-length counting have been known since the 1990s. However, the exponent and fraction cannot be decoded until the regime size is found, since the location of those bit fields depends on it. That reduces the amount of concurrency in the decoding circuit and thus increases worst-case operation latency.

\subsection{A Breakthrough: Bounded Regime Size} 

This leads us to the third step in the history of the choice of posit exponent sizes, which was published in late 2024 \cite{ref3}. The idea is to set a maximum regime size \textit{rS}; that has the effect of reducing the dynamic range, but that effect can be compensated by increasing \textit{eS}. If we choose $\textit{rS}=6$ and want to be able to perform $n$-body astronomical calculations or quantum mechanical calculations involving Planck's constant (about $6.6\times 10^{-34}$), then $\textit{eS}=5$ suffices. For AI, signal processing, and many HPC codes, a smaller \textit{eS} suffices and that increases the bits available for the significand. Both \textit{eS} and \textit{rS} are parameterized in our open-source b-posit design.

\textit{Normal} floats retain an advantage over posits in hardware decode-encode, because fixed field widths enable immediate concurrency in decoding both the exponent and the fraction. For \textit{subnormal} floats, decoding requires counting of leading zeros in the same way that regime decoding does. A format that allows the highly parallel decoding of IEEE 754 floats while preserving the simplicity and superior mathematical properties of posits would offer significant advantages for real-number representation in computer arithmetic. Constraining the size of the posit regime field reduces the number of possible regime-significand size combinations. The regime sequence may still be terminated by the first opposite bit; however, it can also be terminated upon reaching a predefined maximum regime size, \textit{rS}. Hunhold was perhaps the first to observe that limiting the number of regime bits reduces the dynamic range; however, it also makes the decode-encode logic more hardware-friendly and parallel \cite{ref14}. It can also ensure a minimum number of significant bits, which is important for numerical analysis theorems, and unlike both standard posits and IEEE subnormal floats for which all significance is lost at extreme magnitudes. This is highlighted in Fig.~5a and Fig.~5b, where both cases---regime terminated by opposite bit and regime terminated by reaching maximum size---are depicted.

\begin{figure}[ht]
  \centering

  \begin{subfigure}{\linewidth}
    \centering
    \includegraphics[width=\linewidth,valign=t]{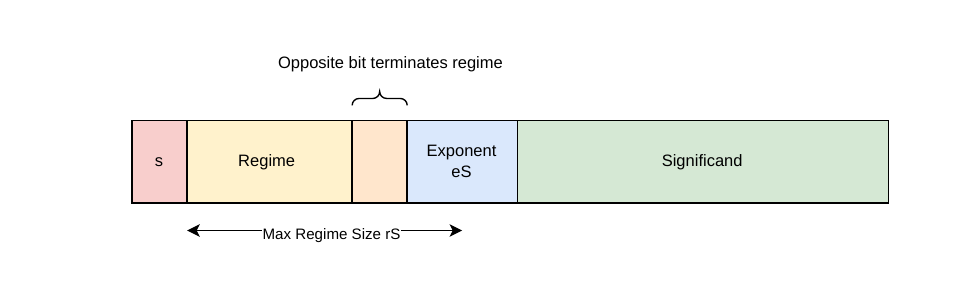}
    \caption{Regime terminated by opposite bit}
    \label{fig:regimeopposite}
  \end{subfigure}

  \vspace{0.3cm}

  \begin{subfigure}{\linewidth}
    \centering
    \includegraphics[width=\linewidth,valign=t]{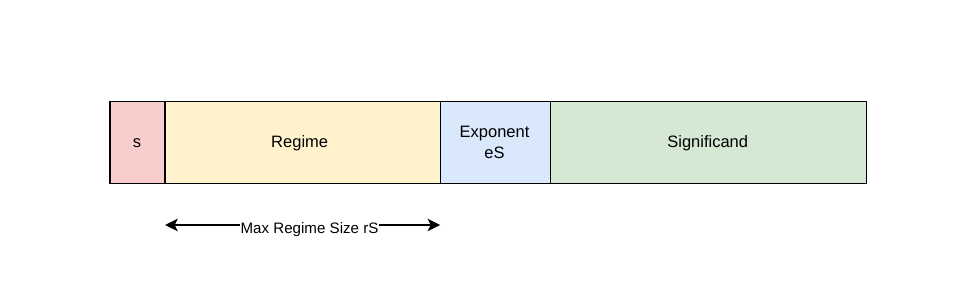}
    \caption{Regime terminated by reaching maximum size \textit{rS}}
    \label{fig:regimemaxedoout}
  \end{subfigure}

  \caption{Illustration of regime termination in the b-posit format.}
  \label{fig:regimeterminations}
\end{figure}

A standard $n$-bit posit has a maximum regime size \textit{rS} equal to $n-1$, and as a result, the relative accuracy plot will exhibit $n-1$ accuracy levels. This is illustrated in Fig.~6a, which depicts the accuracy plot of $\langle 16,2 \rangle$ posits. If \textit{rS} is limited to $6$~bits, as in Fig.~6b, the dynamic range is halved and the tapered accuracy of the plot is flattened. Increasing the exponent size to $eS=3$ compensates for the loss in dynamic range. As shown in Fig.~5, at the cost of $0.3$ decimals of accuracy in the maximum accuracy region, the accuracy does not drop to zero decimals at extreme magnitudes; it never drops below two decimals of accuracy. The parameters \textit{rS} and $eS$ can be tuned to achieve a desired trade-off between relative accuracy (significant digits) and dynamic range for a particular technical workload.

\begin{figure}[ht]
  \centering

  \begin{subfigure}{0.9\linewidth}
    \centering
    \includegraphics[width=\linewidth,valign=t]{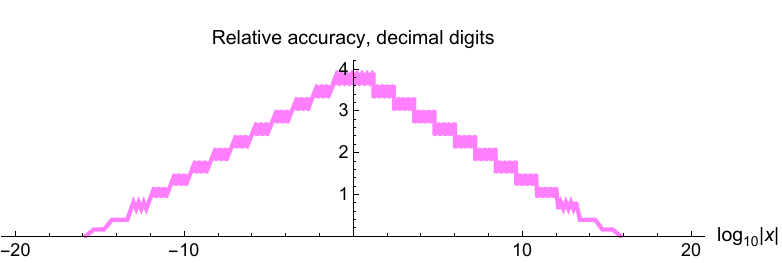}
    \caption{Accuracy Plot for 16-bit standard posit ($\textit{eS}=2$)}
    \label{fig:accplotposit}
  \end{subfigure}

  \vspace{0.3cm}

  \begin{subfigure}{0.9\linewidth}
    \centering
    \includegraphics[width=\linewidth,valign=t]{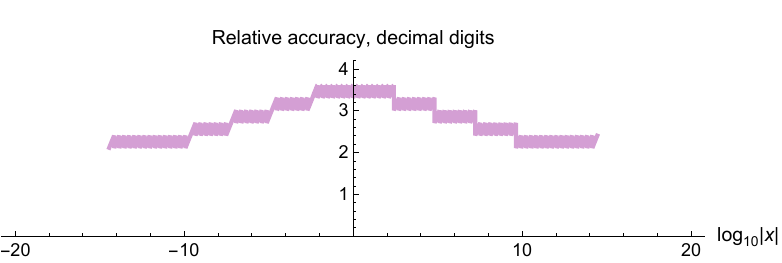}
    \caption{Accuracy plot for 16-bit b-posit ($\textit{rS}=6$ and $\textit{eS}=3$)}
    \label{fig:accplotbposit}
  \end{subfigure}
  \caption{Accuracy plot comparison for standard and bounded posits (16-bit)}
  \label{fig:regime}
  
\end{figure}

A bounded posit, or \emph{b-posit}, is defined as a posit with a maximum regime field size \textit{rS} that is less than $n-1$. As highlighted in~\cite{ref3}, this configuration provides a better match than standard posits to the distribution of real numbers used in AI and HPC workloads; this work shows that it also simplifies the hardware implementation of decode–encode. By constraining \textit{rS} to $6$, the size of the regime is limited to values between $2$ and $6$, resulting in only five possible combinations of regime-significand size. This bounded number of regime sizes enables decode–encode operations to be efficiently realized using basic multiplexers for each of the constituent b-posit fields, with each multiplexer input tapping different parts of the b-posit word.  

This constraint further allows the fields to be decoded in parallel rather than sequentially, thereby significantly reducing latency. In contrast, conventional posit decode–encode hardware relies on components such as shifters and leading-bit detectors that must be applied in sequence, limiting parallelism and increasing area and power consumption. Floating-point subnormal handling hardware similarly requires adders, shifters, and leading-bit detectors. This has led many hardware designers to build noncompliant arithmetic that ignores the subnormal requirement of IEEE Std 754, especially in the design of Graphics Processing Units (GPUs), where the values of subnormal floats are flushed to zero. This means that floating-point bit patterns no longer mean the same when moved from one computer to another, just like the days before there was IEEE Std 754. It also means that $x-y$ can equal zero, even when $x\neq y$.

The hardware b-posit decode-encode demonstrates superior scalability with increasing precision compared to the posit and floating-point implementations, maintaining near-constant latency across word widths from $16$ to $64$ bits. It achieves notable improvements over conventional posit decode–encode, particularly at higher precisions, while also performing competitively against floating-point hardware.

The `brain float' or \textit{bfloat} format introduced in 2018 by Google for AI workloads is another example of a format with a bounded dynamic range \cite{ref13}. Its exponent is like IEEE float32 with an 8-bit exponent, but the fraction length can be 8, 16,$\dots$, 56 in Google's Tensor Processing Unit (TPU). The fixed exponent field means that all bfloats have a dynamic range of $2^{-126}$ to almost $2^{128}$, or about $10^{-38}$ to $~3\times 10^{38}$, independent of precision. The 16-bit version (bfloat16) has been supported by suppliers other than Google, creating a kind of \textit{de facto} industry standard.

A third and more recent format using the bounded-range idea is the \textit{takum} format developed by Laslo Hunhold \cite{ref14}. Takums fit within the posit framework in the way they map 2's complement integers to the projective reals, but they use an additional field to express scaling from $2^{-254}$ to $2^{254}$ (about $10^{-77}$ to $10^{77}$), using from 4 to 11 bits. Like b-posits, they seek to cover a particular dynamic range with a set maximum number of bits of overhead for the scaling, guaranteeing that all precisions greater than that maximum are fraction bits that increase relative accuracy.

Fig.~7 compares four 32-bit formats: IEEE float32 (green), standard posit32 (magenta), takum32 (gray), and the proposed b-posit32 with $\textit{rS}=6, \textit{eS}=5$. Note that all three formats use some form of tapered precision; for floats, the tapering occurs only on the left and exhibits a steep drop in precision. The proposed b-posit32 has a distribution very close to what is needed for general technical computing. For 32-bit precision, both IEEE and the standard posit have a dynamic range too small for, say, Einstein's cosmological constant $\mathrm{\Lambda}=1.4657\times 10^{-52}\mathrm{~m}^{-2}$. The b-posit can represent that value with an accuracy of eight decimal places ($\mathrm{\Lambda}\approx1.4657003\times 10^{-52}$) despite its extreme magnitude.

\begin{figure}[ht]
    \centering
    \includegraphics[width=1\linewidth]{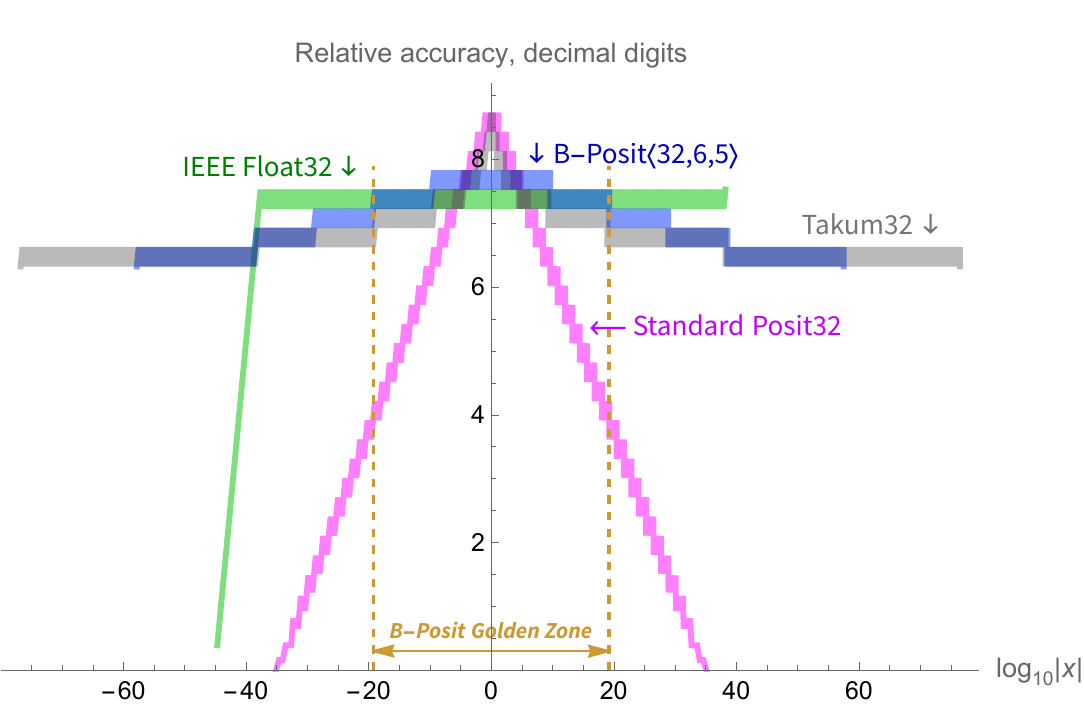}
    \caption{Accuracy plots comparing \texttt{float32}, \texttt{posit32}, and \texttt{b-posit32}}
    \label{fig:accplots}
\end{figure}

The term \emph{Golden Zone} was coined by Florent de Dinechin to denote the region where posits offer at least as much accuracy as floats. For standard 32-bit posits, the Golden Zone extends from $2^{-20}$ to $2^{20}$ (approximately $10^{-6}$ to $10^6$), while b-posit32 extends this region from $2^{-64}$ to $2^{64}$ (approximately $10^{-20}$ to $10^{20}$), and 75\% of the bit patterns fall within that region.

The \emph{fovea} is defined as the region of maximum relative accuracy, that is, the highest level of the tent-shaped accuracy plot. For standard posits, it ranges from $1/16$ to $16$ for any precision $n$. For posit32, the fovea provides four additional bits of significand compared to IEEE floats (a $16\times$ improvement in accuracy). In contrast, for b-posit32, the fovea is massively widened to cover $2^{-32}$ to $2^{32}$ (about $2\times10^{-10}$ to $4\times10^9$), while still delivering twice the accuracy of IEEE floats in that region.

Only experiment will determine which distribution will produce the best answers, and it will certainly depend on the application. For the required-value distributions studied so far, a bell curve seems to be by far the most common case, and the b-posit produces an excellent approximation to that distribution. The takum format distribution is the reverse of a bell curve, with a sharp point in the center and concave-up regions to either side. We plan extensive experiments to compare b-posits and takums.

This work presents the first fully parameterizable and scalable b-posit decoder and encoder architectures, with a focus on parallel bit-field processing to minimize delay. The paper is organized as follows: Section 2 reviews the prior art, outlining existing posit decode–encode methodologies and techniques for subnormal handling and exception detection in floating-point. Section 3 details the design of the proposed b-posit decoder and encoder circuits. Section 4 describes the post-layout timing results of the proposed circuits, benchmarking them against posit and floating-point implementations across various input vectors and technology nodes. Finally, Section 5 concludes the paper.

\section{Prior Art}
\subsection{Floating-Point Arithmetic}
The IEEE floating-point standard specifies five classes of exceptions---\emph{invalid operation}, \emph{division by infinity}, \emph{overflow}, \emph{underflow}, and \emph{inexact}. In addition to these exceptions, the standard requires support for special cases known as \emph{subnormal numbers}, sometimes called ``gradual underflow''. Unlike exceptions, subnormals employ a distinct representation: they are defined as numbers with an exponent field equal to zero while retaining a nonzero significand. In this representation, the implicit leading one is omitted, thereby allowing for leading zeros. Arithmetic involving subnormal operands requires a normalization procedure, wherein the leading one in the significand is detected, followed by an appropriate left shift of the significand and a corresponding adjustment of the exponent which must be augmented with an additional significant bit to prevent underflow. 

Most floating-point hardware implementations in the literature do not explicitly account for the cost of exception detection and subnormal handling. As a result, floating-point arithmetic is often perceived to be significantly less expensive than posit arithmetic, which inherently incorporates both decode and encode stages in addition to the arithmetic operation. Berkeley's HardFloat~\cite{ref4} is one implementation that explicitly includes hardware support for exception detection and subnormal handling. The HardFloat library performs floating-point operations in three stages:  
\begin{itemize}  
    \item \textbf{Decode:} The input operand, expressed in IEEE-754 format, is translated into HardFloat's internal representation. This step includes detection of Not-a-Number (NaN) and signed infinity exception values, and normalization of subnormal values for uniform processing within the arithmetic units. It also involves turning the biased exponent into a 2's complement integer, and prepending the fraction bits with the implied leading bit (\texttt{0} for subnormal floats and \texttt{1} for normal floats).
    \item \textbf{Arithmetic:} The internal representation is processed by the arithmetic units to execute the specified operation. During this stage, exception flags are generated to signal conditions such as invalid operation, overflow, underflow, or inexact results.  
    \item \textbf{Encode:} The computed result in HardFloat form is translated back into IEEE-754 representation. This conversion accounts for the exceptions raised in the arithmetic stage and appropriately encodes subnormal numbers and exceptions. The signed exponent is converted back to a biased exponent, and the implied first bit of the significand is stripped to form the fraction. The final step is to apply rounding, according to the rounding mode indicated by the processor flags.

\end{itemize}  
\begin{figure}[ht]
    \centering
    \includegraphics[width=0.8\linewidth]{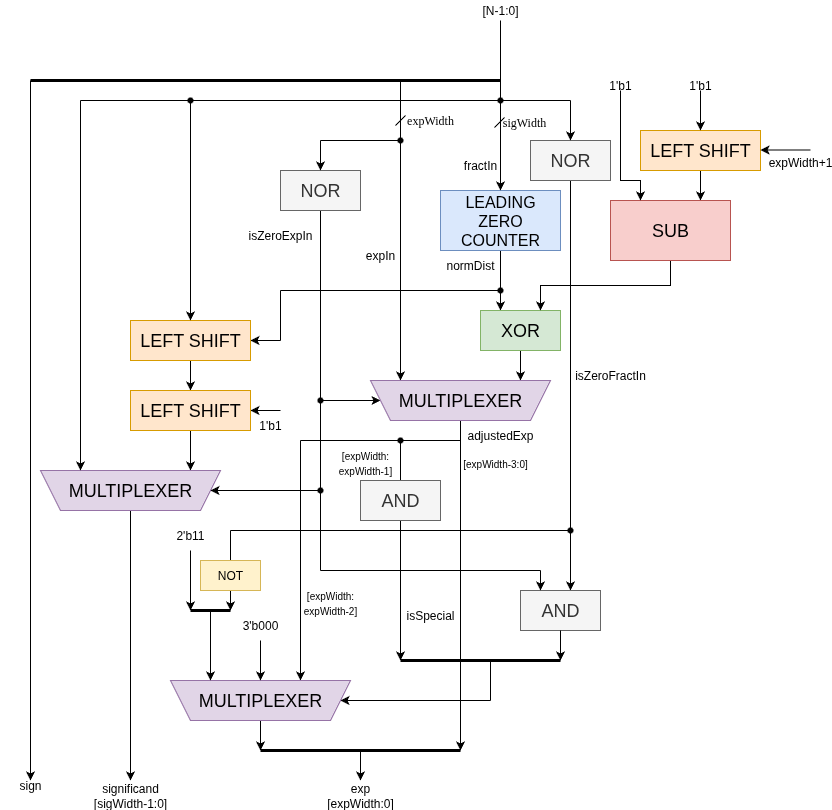}
    \caption{Floating-Point Decoder}
    \label{fig:floatdecoder}
\end{figure}
The floating-point decode-encode is analogous to the decode-encode of its posit and b-posit counterparts. The floating-point decode hardware is depicted in Fig.~8. HardFloat introduces a recoded format that represents the same set of values as the IEEE-754 standard, but in a slightly modified encoding. Similarly to the standard representation, the recoded format consists of sign, exponent, and fraction fields, with the distinction that the exponent is allocated an additional bit. For example, the standard 32-bit single-precision format is recoded into a 33-bit representation, comprising 1 sign bit, 9 exponent bits, and 23 fraction bits. This additional exponent bit enables a more uniform treatment of special cases. In particular, the recoded representation simplifies the hardware implementation of floating-point arithmetic by allowing subnormal values to be internally normalized, thereby enabling them to be processed in a manner consistent with normalized numbers.

\begin{figure}[ht]
    \centering
    \includegraphics[width=0.8\linewidth]{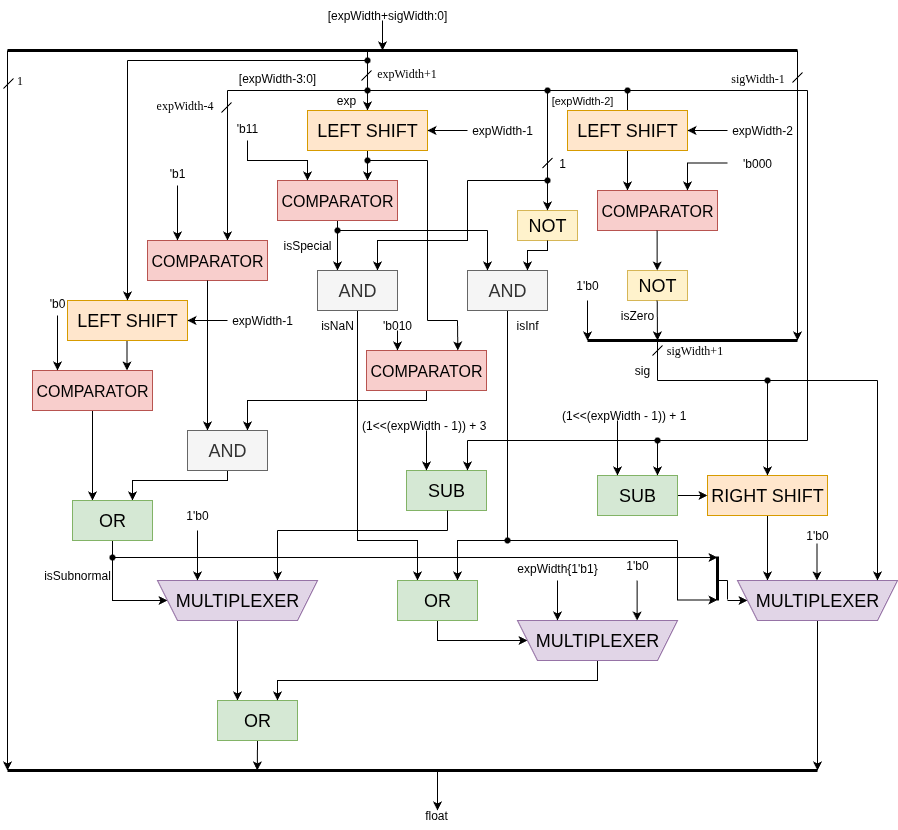}
    \caption{Floating-Point Encoder}
    \label{fig:floatencoder}
\end{figure}

Once the arithmetic process is complete, the HardFloat result is converted back to the standard IEEE-754 form. The input, expressed in the recoded representation with an additional exponent bit, is first decoded into its constituent fields: sign, signed exponent, and significand, along with classification flags identifying NaN, infinity, zero, or subnormal cases. A dedicated logic block determines whether the input corresponds to a subnormal value and, if so, computes the required right-shift distance for proper encoding. The exponent field is then generated by mapping normalized values to biased exponents, forcing all \texttt{1} bits for NaN and infinity and all \texttt{0} bits for subnormals. Similarly, the fraction field is formed by right-shifting the significand for subnormal numbers, zeroing it in the case of infinity, or directly passing it through for normalized values. Finally, the sign, exponent, and fraction fields are concatenated to produce a compliant IEEE-754 output representation. The floating-point encode process is depicted in Fig.~9, showing all steps except the final rounding according to the rounding mode flags.

Overall, the floating-point decode-encode pipeline is analogous to the corresponding stages in posit and b-posit arithmetic. By introducing a recoded format with an additional exponent bit, HardFloat achieves a more uniform treatment of special cases, particularly subnormal numbers, thereby simplifying hardware implementation and ensuring consistent arithmetic processing.  

\subsection{Standard Posit Arithmetic}
Similarly to floats, operand fields must first be decoded, and after computation, the result must be re-encoded into the original format. A key difference is that there are no ``subnormal posits'' to check for; whereas floats must support subnormals so that $x-y=0$ cannot happen if $x\neq y$, that mistake cannot occur with the posit format.

\begin{figure}[ht]
    \centering
    \includegraphics[width=0.8\linewidth]{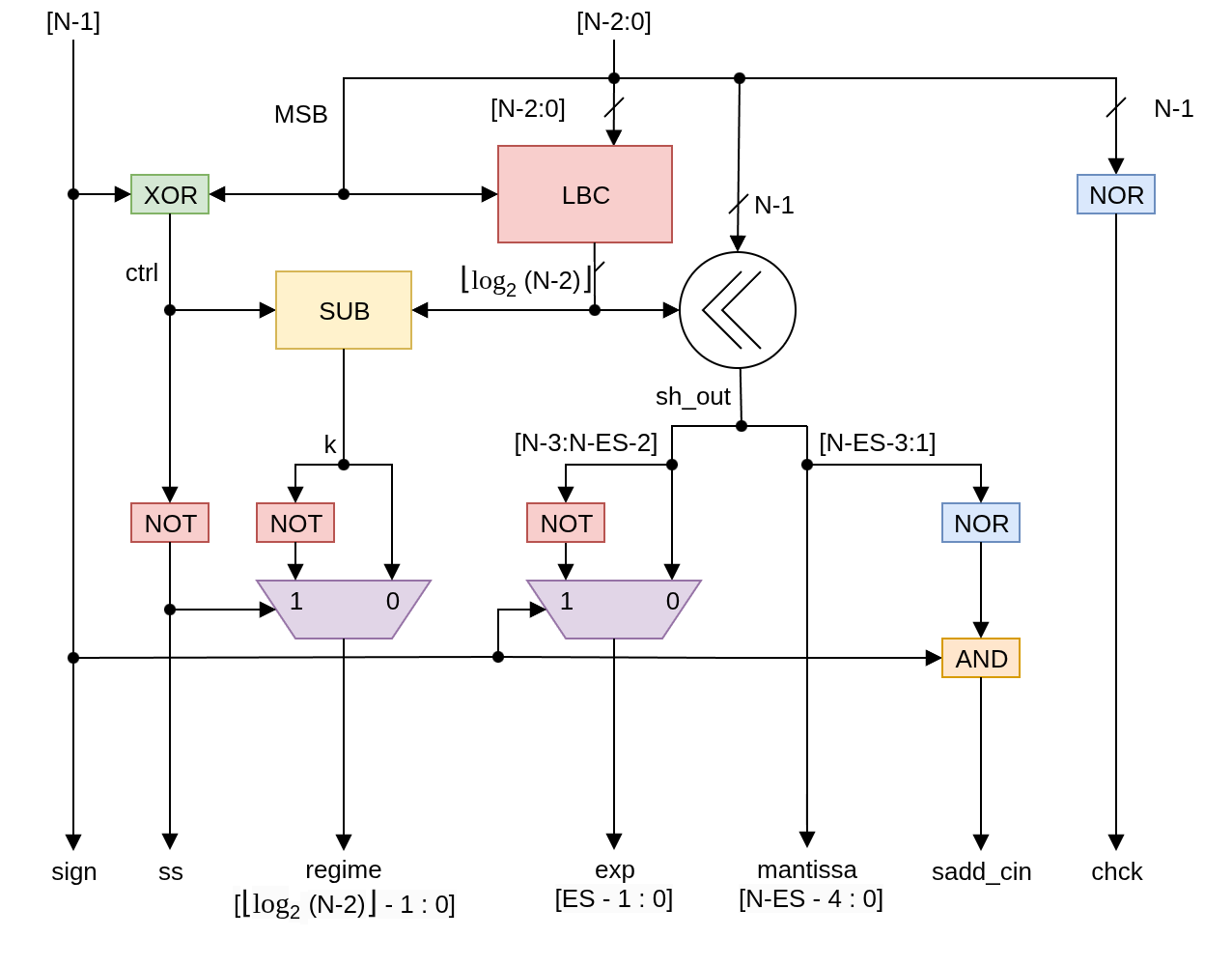}
    \caption{Posit Decoder}
    \label{fig:positdecoder}
\end{figure}

A key step is to extract the regime, exponent, and fraction of each operand. As stated above, the regime’s variable length prevents parallel decoding of the significand, because the exponent and fraction cannot be identified until the regime length has been determined. As a result, posit arithmetic follows the following steps:
\begin{itemize}
  \item \textbf{Posit Decode} — The input operand is translated into an intermediate float-like format. The value of the regime, which is encoded by its length, is first unpacked as a signed integer value, after which the exponent and fraction are extracted.
  \item \textbf{Arithmetic} — The intermediate float-like representation undergoes the specified arithmetic operation and rounding in the same manner as floating-point arithmetic.
  \item \textbf{Posit Encode} — At the end of every operation, the resulting regime, exponent, and fraction are packed into the corresponding posit format. Rounding is applied, and there is only one rounding mode (round-to-nearest, ties-to-even, the default mode of IEEE Std 754).
\end{itemize}

\begin{figure}[ht]
    \centering
    \includegraphics[width=0.8\linewidth]{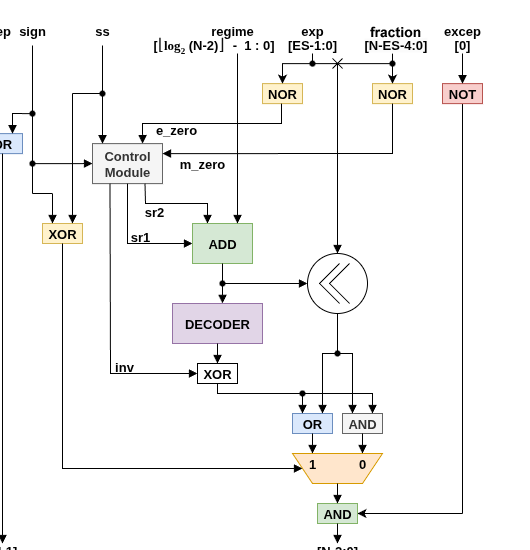}
    \caption{Posit Encoder}
    \label{fig:positencoder}
\end{figure}

The posit decoder proposed in \cite{ref6} is highlighted in Fig.~10 (please excuse the use of the denigrated term ``mantissa'' for the fraction; the term is still in common use, although it creates confusion). In a posit decode process, the width of the regime field is determined by a Leading Bit Counter (LBC). Once the width of the regime field is computed, it is sent as input to a left-shifter from which the exponent and significand are derived. A simple reduction NOR operation derives the ``chck'' bit as an exception detector, which is set when the input posit is either zero or NaR. Other posit decoder implementations follow similar steps with minor differences in the method of extracting the regime field width \cite{ref7,ref8,ref9,ref10,ref11,ref12}.

The posit encoder circuit proposed in \cite{ref6} is depicted in Fig.~11. The encoder packs the sign, regime, exponent, and fraction into posit form in a process opposite to the decode. The regime value is determined from the input regime string, which is then used to appropriately shift the significand.

\section{Proposed B-Posit Decoder and Encoder Circuits}
\subsection{Proposed B-Posit Decoder}
We notate b-posits as a tuple $\langle N,\textit{rS},\textit{eS}\rangle$ for precision $N$, maximum regime size \textit{rS}, and exponent size \textit{eS}. The proposed b-posit uses $\textit{rS}=6$ bits, thereby limiting the number of possible regime sizes to five (ranging from 2 to 6). This is in contrast with the conventional \(\langle N,\textit{eS} \rangle\) posit, where the regime field may assume up to \(N-1\) width. The decoding process always begins with identifying the size of the regime field. For this purpose, the first five bits, including the sign bit and the most significant regime bit, are XORed with the regime MSB. This simplifies the detection process by recognizing a sequence of \texttt{0} s terminated by a \texttt{1} or by reaching the maximum allowable size \textit{rS}. Using five bits ensures correct operation even in the extreme case where the regime occupies its full six-bit width. These five bits are then mapped to a one-hot encoded vector, as summarized in Table 2 using simple combinational logic comprising AND, NOT gates. A \textit{one-hot} encoded string is a binary representation where exactly one bit is ‘1’ and all the others are ‘0’.

\begin{table}[h]
\centering
\begin{tabularx}{\textwidth}{|*{5}{>{\centering\arraybackslash}X|}*{6}{>{\centering\arraybackslash}X|}}
\hline
\multicolumn{5}{|c|}{Input[N-3:N-7] after XOR} &
\multicolumn{6}{c|}{One-Hot String} \\
\hline
1 & X & X & X & X & 1 & 0 & 0 & 0 & 0 & 0 \\
\hline
0 & 1 & X & X & X & 0 & 1 & 0 & 0 & 0 & 0 \\
\hline
0 & 0 & 1 & X & X & 0 & 0 & 1 & 0 & 0 & 0 \\
\hline
0 & 0 & 0 & 1 & X & 0 & 0 & 0 & 1 & 0 & 0 \\
\hline
0 & 0 & 0 & 0 & 1 & 0 & 0 & 0 & 0 & 1 & 0 \\
\hline
0 & 0 & 0 & 0 & 0 & 0 & 0 & 0 & 0 & 0 & 1 \\
\hline

\end{tabularx}
\caption{Regime size as one-hot string}
\label{tab:onehot-bposit}
\end{table}

The one-hot encoded string represents all the possible sizes of the regime field in a unique pattern. This one-hot string is used as the select input to a common multiplexer for the exponent and regime fields, each input to the multiplexer tapping different substrings of the b-posit as shown in Fig.~12. For example, when the regime field of the b-posit is \texttt{01}, the smallest possible size, the one-hot encoded string is computed as \texttt{100000}. The input corresponding to this value of the one-hot encoded string is the b-posit excluding the sign bit and the 2 regime bits. Similarly, when the regime field is \texttt{001}, the one-hot encoded string is \texttt{010000}, and the corresponding input is the b-posit excluding the sign bit and 3 regime bits, appended with 0 to match the width. At the output of the multiplexer, the most significant $eS$ bits are taken as the raw exponent bits, and the remaining bits form the significand. Since the b-posit is packed in signed form, a 2's complement is needed to be performed to determine the various fields of the b-posit. 

Consequently, the raw exponent is XORed with the sign bit to generate the final exponent field. This effectively performs a 1's-complement operation on the raw exponent. The approach is valid only when at least one significand bit is set to \texttt{1}. The underlying rationale is that in a 2's-complement operation, bits remain unchanged from the least-significant position up to the first `1', after which all bits are inverted. The signal \texttt{exp\_cin} accounts for cases where the significand is zero and the sign is negative, essentially transforming the 1's complement to the 2's complement when necessary. This term can be added to the exponent during the arithmetic stage, rather than the decode stage, since the significand computation dominates the timing critical path. As a result, the latency of the 2's-complement operation is effectively reduced to a single XOR operation, while the significand remains in signed form.

\begin{figure}[ht]
    \centering
    \includegraphics[width=\linewidth]{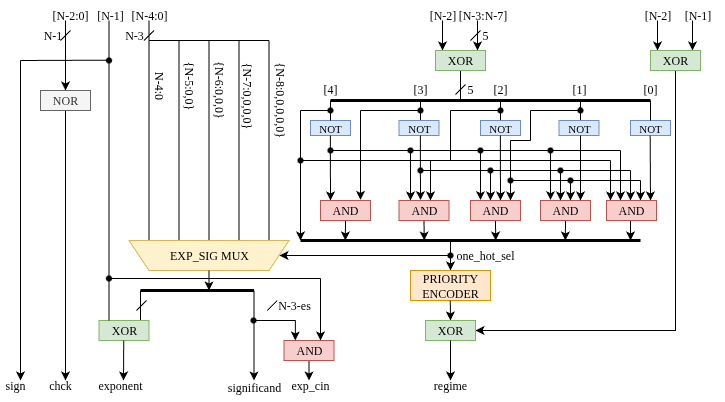}
    \caption{Proposed B-Posit Decoder}
    \label{fig:bpositdecoder}
\end{figure}

The raw regime is derived from a priority encoder with the one-hot encoded string as input as shown in Table~2. The priority encoder takes multiple input signals and outputs a binary code representing the highest priority active input. The entire decode process is accomplished in two simple steps: simple combinational logic to deduce the regime size, followed by a multiplexer with inputs as different tap points of the b-posit based on the regime size and a priority encoder to determine the regime value. Since both the multiplexer and the priority encoder work in parallel and take input as the one-hot encoded string from Table ~2, the critical path delay of the entire b-posit decoder is- XOR, NOT, AND, priority encoder/multiplexer. As the b-posit size increases, the combinational logic used to determine the regime field width, as well as the priority encoder, remain unchanged since the regime length is still limited to a maximum of \textit{6} bits. The multiplexer also remains a 5-input structure, with only the width of its signal inputs increasing. The delay of a multiplexer primarily depends on the number of inputs, since a greater fan-in increases logic depth. When only the input width changes, the effect is limited to higher power and area consumption, as this can be visualized as multiple single-bit multiplexers operating in parallel. Consequently, while power and area scale proportionally with increasing precision, the overall delay remains nearly constant across different precisions.

This contrasts with posit decoders in the literature where the critical path generally consists of leading-bit detectors, shifters, multiplexers, and adders, all operating \emph{sequentially}. The logic depth and overall complexity of the leading-bit detector grow significantly as the input size increases, mainly because the detection process involves a hierarchical network of comparators or priority encoders that must propagate through more logic levels to identify the most significant ‘\textit{1} bit. As the number of input bits increases, the number of potential leading-bit positions also increases, thereby increasing both the gate count and the propagation delay associated with the detection logic. Similarly, the barrel shifter is typically realized as a collection of multiplexers operating in parallel scales with precision. Each bit of the output requires a dedicated multiplexer chain capable of selecting among all possible shift positions. Consequently, as the precision increases, both the number of multiplexers and the number of inputs per multiplexer increase linearly. Hence, for high-precision implementations, the leading-bit detector and shifter together can become major contributors to the critical path and overall hardware complexity.

The proposed design exhibits an almost constant delay across different precisions, as the critical combinational paths remain largely unaffected by the increase in bit width. The only observable overheads arise in terms of area and power consumption, both of which scale proportionally with precision. This property stands in contrast to conventional posit and floating-point decode architectures, where the delay tends to increase with operand width due to the presence of more complex combinational logic and critical delay paths. The b-posit decode methodology cannot be directly applied to standard posit decoding, as the regime field in conventional posits can extend across nearly the entire posit width. This characteristic makes direct hardware realization highly inefficient. For instance, implementing a 64-bit posit decoder would require a multiplexer with up to 63 inputs to accommodate the full range of possible regime lengths. Such a design would result in considerable increases in both area utilization and dynamic power consumption, making it unsuitable for high-precision implementations.

In contrast, the b-posit format enforces a strict upper bound on the regime field length, restricting it to a maximum of five bits. As a result, the associated multiplexer remains a compact 5-input structure regardless of precision. This bounded design ensures that the critical path delay remains nearly constant even as the bit width increases. The inherent structural simplicity of the b-posit architecture therefore enables superior scalability and energy efficiency while maintaining consistent performance across varying precision levels.

\subsection{Proposed B-Posit Encoder}
The b-posit decoder translates the regime bit string into a \textit{4}-bit regime value while simultaneously identifying the exponent and mantissa fields of the b-posit. The proposed b-posit encoder follows a similar approach, leveraging the bounded regime length to enable the parallel packing of all constituent fields into the final posit representation. The encoding process begins by determining the width of the regime field, which is derived directly from the computed regime value. To achieve this, the three least significant bits of the regime value are XORed with its most significant bit (MSB). This operation is necessary because both the regime value and its 1’s-complement counterpart produce regime fields of equal width when encoded, as illustrated in Table~3. The resulting XOR output serves as the multiplexer select signal, which identifies one of five possible packing configurations, corresponding to regime field lengths of 2, 3, 4, 5, or 6 bits.
\begin{table*}[ht]
\centering
\label{tab:regime_size}
\begin{tabularx}{\textwidth}{|>{\centering\arraybackslash}X
                              |>{\centering\arraybackslash}X
                              |>{\centering\arraybackslash}X|}
\hline
\multicolumn{2}{|c|}{\textbf{Regime [3:0]}} & \textbf{Regime Size} \\
\hline
\texttt{0000} & \texttt{1111} & 2 \\
\hline
\texttt{0001} & \texttt{1110} & 3 \\
\hline
\texttt{0010} & \texttt{1101} & 4 \\
\hline
\texttt{0011} & \texttt{1100} & 5 \\
\hline
\texttt{0100} & \texttt{1011} & 6 \\
\hline
\texttt{0101} & \texttt{1010} & 6 \\
\hline
\end{tabularx}
\caption{Regime size determination from the four-bit regime field.}
\end{table*}

For instance, when the regime size is two, the complete exponent and significand fields are packed alongside the two regime bits. For a regime size of three, the three regime bits, together with the exponent and the significand excluding its LSB, are incorporated. As the regime size increases further, the fraction width decreases proportionally, as illustrated in Fig.~13. The regime string is derived from the regime value by using a simple binary decoder circuit. It is important to note that the binary decoder is a combinational circuit that converts a binary value to a one-hot string, this is different from the b-posit decoder discussed in the previous subsection. The three regime bits XORed with the MSB are re-used as input to the decoder. The output of the binary decoder is padded with a \texttt{0} bit as the MSB and is used to derive the regime string as shown in Table ~4. When the regime after 1's complement is \textit{100} or \textit{101}, the size of the regime remains \textit{6} as shown in Table ~4. This is because the regime strings \textit{000000} and \textit{000001} are both \textit{6}-bits wide, however, have different values. The decoder is implemented as a $3\times6$ binary decoder rather than a $3\times8$ binary decoder as the input does not exceed \texttt{100}. 

\begin{figure}[ht]
    \centering
    \includegraphics[width=\linewidth]{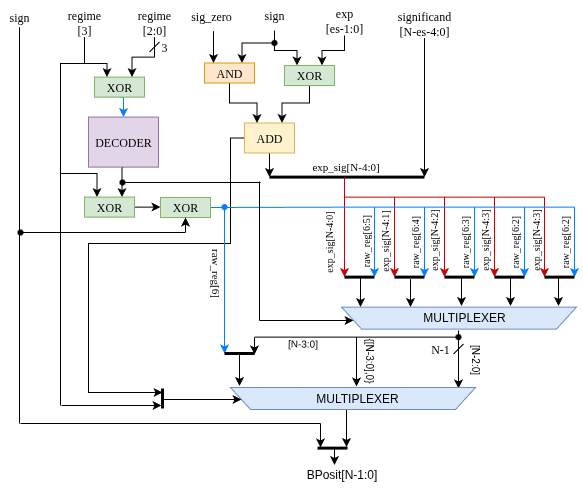}
    \caption{Proposed B-Posit Encoder}
    \label{fig:bpositencoder}
\end{figure}

\begin{table}[h]
\centering
\begin{tabularx}{\textwidth}{|>{\centering\arraybackslash}X
                                |>{\centering\arraybackslash}X
                                |>{\centering\arraybackslash}X
                                |>{\centering\arraybackslash}X|}
\hline
\makecell{Regime after 1's\\ Complement} &
Regime Size &
Decoder Output &
\makecell{Intermediate\\ Regime String} \\
\hline
\texttt{000} & 2 & \texttt{100000} & \textbf{0100000} \\
\hline
\texttt{001} & 3 & \texttt{010000} & \textbf{0010000} \\
\hline
\texttt{010} & 4 & \texttt{001000} & \textbf{0001000} \\
\hline
\texttt{011} & 5 & \texttt{000100} & \textbf{0000100} \\
\hline
\texttt{100} & 6 & \texttt{000010} & \textbf{0000010} \\
\hline
\texttt{101} & 6 & \texttt{000001} & \textbf{0000001} \\
\hline
\end{tabularx}
\caption{Regime string determination from regime value}
\end{table}

The intermediate regime string derived from the decoder depends on the input of the decoder, and the number of bits of the string to use depends on the size of the regime field, as shown in Table ~4. For example, when the 1 complement of the regime yields \texttt{011}, the decoder output, after prepending a \texttt{0} bit, is \texttt{0000100}. Given that the regime size in this case is five, the five most significant bits of the string are extracted as the intermediate regime. The final regime is then obtained through an XOR operation involving the regime MSB and the sign bit, as illustrated in Fig.~13. The exponent is converted into 2’s complement representation via an XOR with the sign bit, followed by an increment in cases where the sign is negative and the fraction is zero. Ultimately, the regime, exponent, and fraction corresponding to each regime size are supplied as inputs to the multiplexer, as highlighted in Fig.~13. Finally, if the exponent field overflows after the 2's complement, instead of performing an addition with the regime field, the change in the final regime string is accounted for using another multiplexer. If the exponent overflows and the regime value is negative, the string size is reduced by one. Similarly, if the exponent overflows and the regime value is positive, the size of the string increases by one.

Similarly to the b-posit decoder, the b-posit encoder demonstrates excellent scalability with increasing precision, particularly in terms of latency. This efficiency arises from the regime field being bounded to 6 bits, ensuring that the logic for regime-width determination in the initial encoding stage remains unchanged across precisions. The critical path of the circuit consists of three XOR gates, one binary decoder, and two multiplexers. In comparison, the posit encoder exhibits a more complex critical path involving a NOR gate, control-module combinational logic, a binary adder, a shifter, a binary decoder, two AND gates, and a multiplexer. The shifter and binary decoder contribute greatly to the increase in logic depth and complexity with higher precision. In contrast, the proposed b-posit encoder maintains scalability because only the multiplexer scales with precision. The multiplexer retains its 5-input configuration, with the input width as the sole variable parameter. As delay primarily depends on the number of inputs, wider inputs impact power and area, but have negligible influence on delay. Consequently, power and area grow proportionally with precision, while latency remains nearly constant. Notably, such a multiplexer-based structure is infeasible in standard posit encoding due to the constraints discussed in the previous b-posit decode subsection.

\section{Results}
The proposed B-Posit decoder, encoder along with IEEE float decode-encode and standard posit decode-encode circuits have been implemented from RTL design to post-layout timing using the SiliconCompiler tool attached with the open source \emph{freepdk45} technology library. The results are highlighted in Tables 5 and 6 and Figures 14,15 and 16.
\begin{figure}[ht]
  \centering

  \begin{subfigure}{0.5\linewidth}
    \centering
    \includegraphics[width=\linewidth,valign=t]{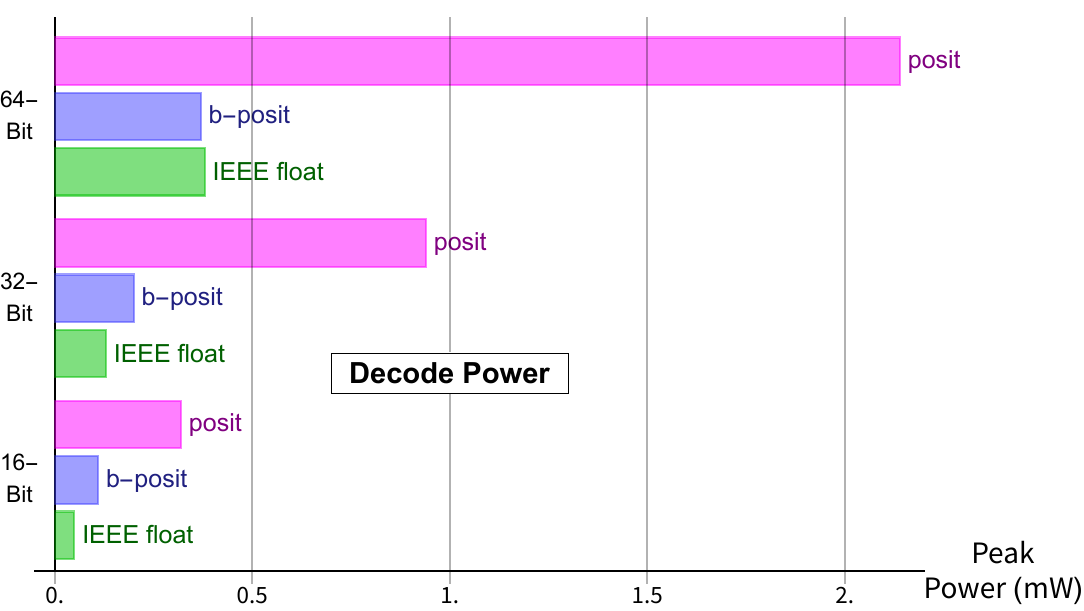}
    \caption{Decode Peak Power (mW)}
    \label{fig:decodepower}
  \end{subfigure}

  \vspace{0.3cm}

  \begin{subfigure}{0.5\linewidth}
    \centering
    \includegraphics[width=\linewidth,valign=t]{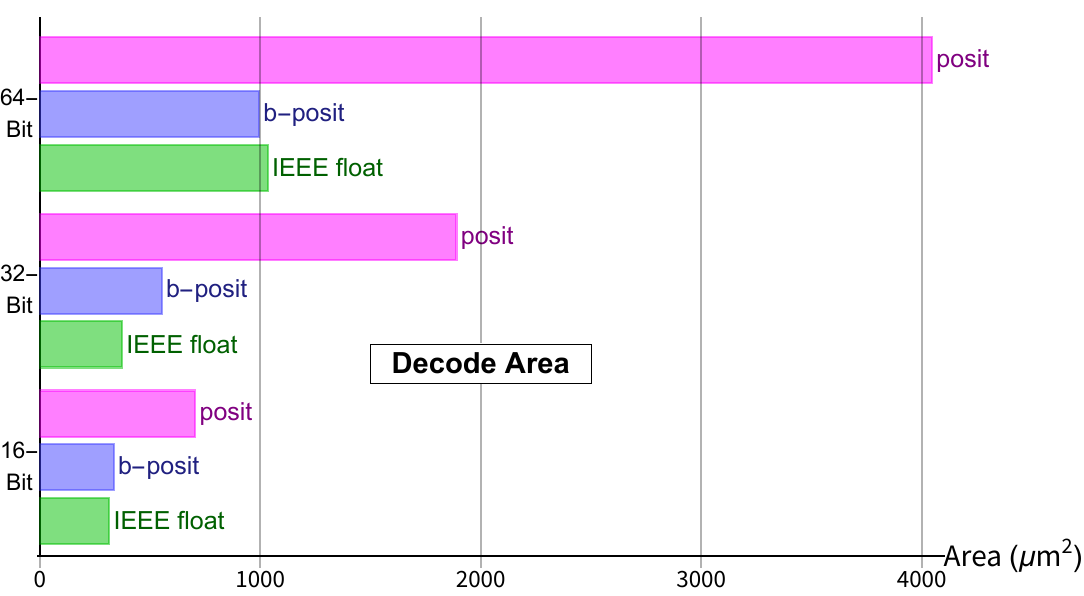}
    \caption{Decode Area ($\mu\mathrm{m}^2$)}
    \label{fig:decodearea}
  \end{subfigure}

  \vspace{0.3cm}

  \begin{subfigure}{0.5\linewidth}
    \centering
    \includegraphics[width=\linewidth,valign=t]{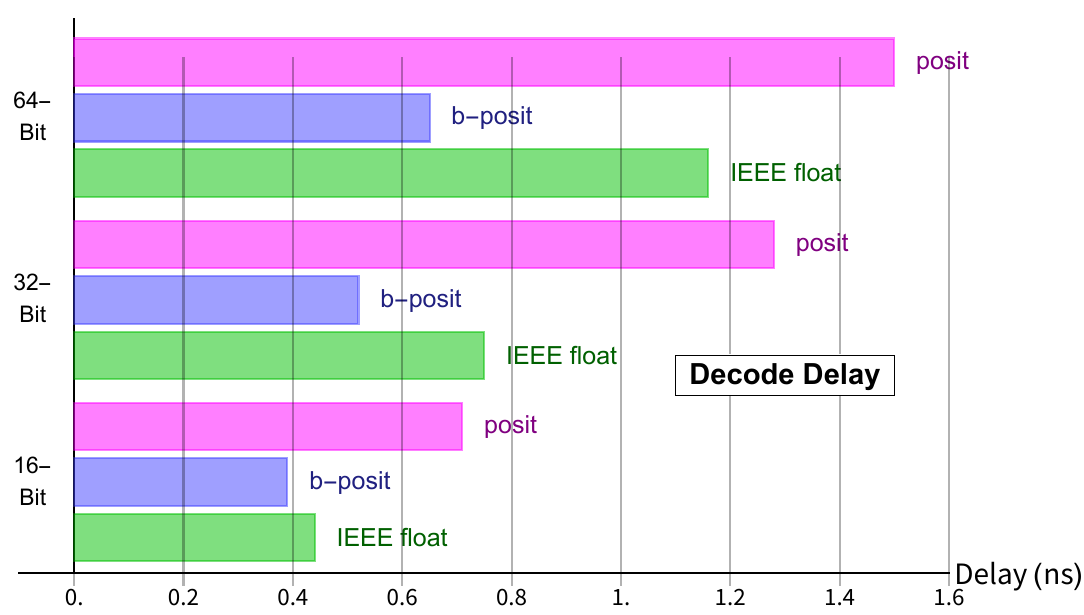}
    \caption{Decode Delay (ns)}
    \label{fig:decodedelay}
  \end{subfigure}
  
  \caption{Decode cost comparisons between posits, b-posits, and floats}
  \label{fig:decodecost}
  
\end{figure}
\begin{table}
\begin{tabular}{|c|c|c|c|c|}
\hline
Configuration & Design                 & Peak Power (mW) & Area($\mu$m$^2$) & Delay(ns) \\ \hline
16            & Floating-Point Decoder & 0.05            & \textbf{~315}                & 0.44      \\ \hline
$\langle16,6,5\rangle$      & B-Posit Decoder        & 0.11             & \textbf{~335}                & 0.39      \\ \hline
$\langle16,2\rangle$        & Posit Decoder          & 0.32            & \textbf{~705}                & 0.71      \\ \hline
32            & Floating-Point Decoder & 0.13            & \textbf{~373}                & 0.75      \\ \hline
$\langle32,6,5\rangle$      & B-Posit Decoder        & 0.20            & \textbf{~553}                & 0.52      \\ \hline
$\langle32,2\rangle$        & Posit Decoder          & 0.94            & \textbf{1890}               & 1.28      \\ \hline
64            & Floating-Point Decoder & 0.38            & \textbf{1034}                        & 1.16      \\ \hline
$\langle64,6,5\rangle$      & B-Posit Decoder        & 0.37            & \textbf{~994}                         & 0.65      \\ \hline
$\langle64,2\rangle$        & Posit Decoder          & 2.14            & \textbf{4047}                        & 1.50       \\ \hline
\end{tabular}
\caption{Comparison between b-posit, posit and floating-point decode at 45 nm}
\end{table}

At 16-bit precision, the b-posit decoder shows a significant improvement in area, power, and delay compared to the standard posit decoder. This includes a 52\% reduction in area, a 45\% reduction in delay, and a 66\% reduction in power. However, it still draws more power and is slower than floating-point decode at this precision. The proposed b-posit decoder scales extremely well in terms of delay compared to the posit and floating-point decode.

For 32-bit results, our 32-bit b-posit decoder circuits achieve significant improvements (79\% less power consumption, 71\% smaller area, and a 60\% reduction in latency) compared to posit decoders. At this 32-bit precision, the delay of the b-posit is 69\% that of the floating-point decode time; in other words, the decoding of the b-posit is \textbf{39\% faster than the IEEE float decode of the same precision}, which we believe is a first for a format within the posit framework. However, there remains a power and an area overhead.

At 64-bit precision, the proposed b-posit decoder significantly outperforms both floating-point decode and posit decode owing to its superior scalability. The b-posit decoder is 
\textbf{more than twice as fast as the floating-point decoder, but it uses slightly \textit{less} power and area}. Compared to standard posit decode, it is almost three times as fast, draws less than one-sixth the power, and is four times smaller. Bar graphs make comparisons more clear (Fig. 14).

\begin{figure}[ht]
  \centering

  \begin{subfigure}{0.5\linewidth}
    \centering
    \includegraphics[width=\linewidth,valign=t]{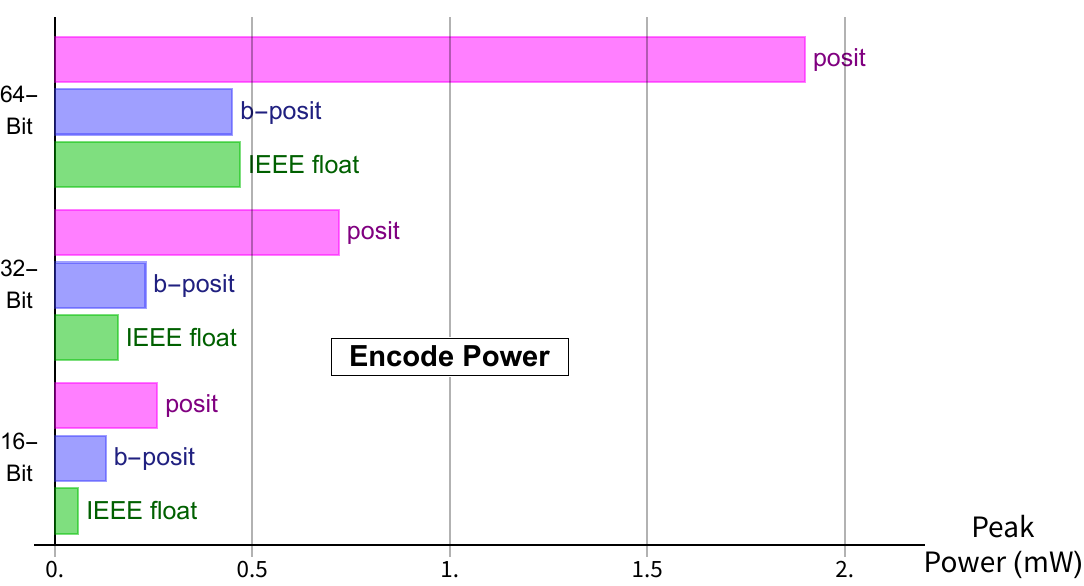}
    \caption{Encode Peak Power (mW)}
    \label{fig:encodepower}
  \end{subfigure}

  \vspace{0.3cm}

  \begin{subfigure}{0.5\linewidth}
    \centering
    \includegraphics[width=\linewidth,valign=t]{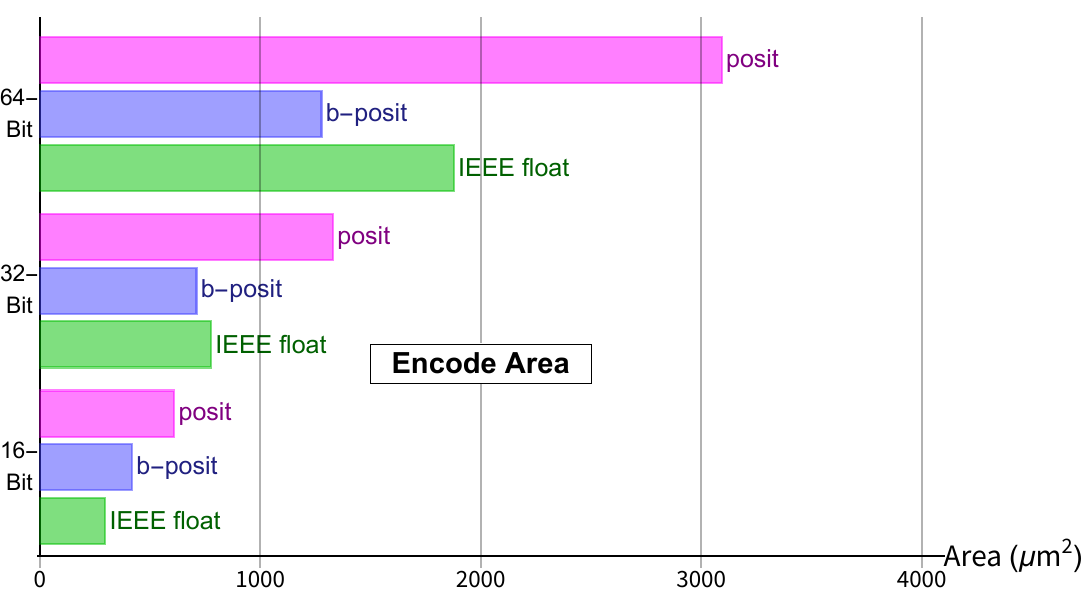}
    \caption{Encode Area ($\mu\mathrm{m}^2$)}
    \label{fig:encodearea}
  \end{subfigure}

  \vspace{0.3cm}

  \begin{subfigure}{0.5\linewidth}
    \centering
    \includegraphics[width=\linewidth,valign=t]{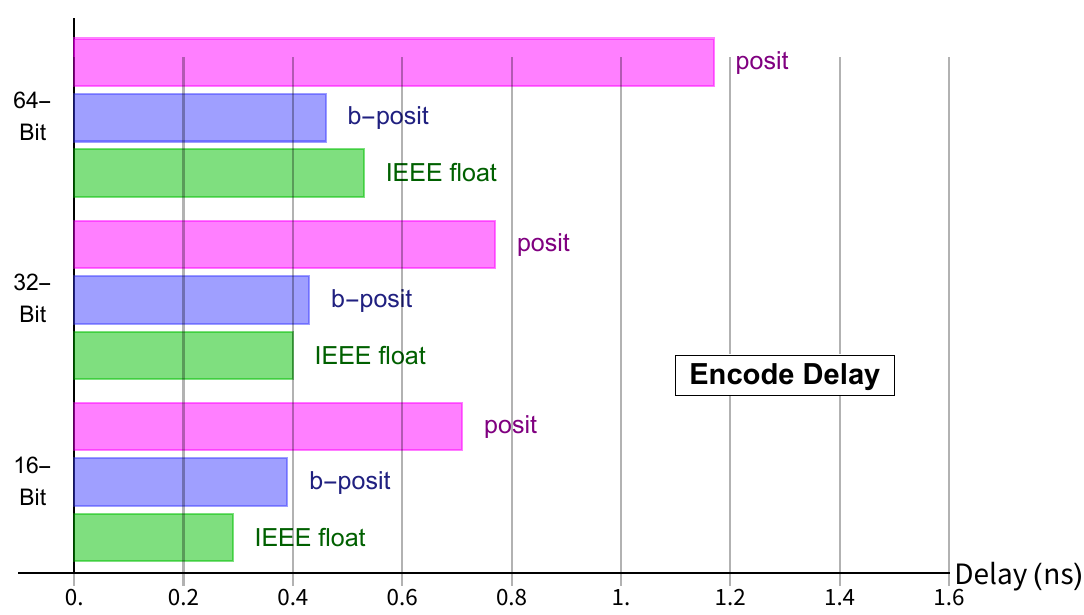}
    \caption{Encode Delay (ns)}
    \label{fig:encodedelay}
  \end{subfigure}
  
  \caption{Encode cost comparisons between posits, b-posits, and floats}
  \label{fig:encodecost}
  
\end{figure}

\begin{table}
\begin{tabular}{|c|c|c|c|c|}
\hline
Configuration & Design                 & Peak Power (mW) & Area($\mu$m$^2$) & Delay(ns) \\ \hline
16            & Floating-Point Encoder & 0.06            & \textbf{297}                & 0.29      \\ \hline
$\langle16,6,5\rangle$      & B-Posit Encoder        & 0.13            & \textbf{418}                & 0.39      \\ \hline
$\langle16,2\rangle$        & Posit Encoder          & 0.26            & \textbf{610}                & 0.71      \\ \hline
32            & Floating-Point Encoder & 0.16            & \textbf{777}                & 0.4       \\ \hline
$\langle32,6,5\rangle$      & B-Posit Encoder        & 0.23            & \textbf{711}                & 0.43      \\ \hline
$\langle32,2\rangle$        & Posit Encoder          & 0.72            & \textbf{1330}               & 0.77      \\ \hline
64           & Floating-Point Encoder & 0.47            & \textbf{1878}                        & 0.53      \\ \hline
$\langle64,6,5\rangle$      & B-Posit Encoder        & 0.45            & \textbf{1278}                        & 0.46       \\ \hline
$\langle64,2\rangle$        & Posit Encoder          & 1.9             & \textbf{3093}                        & 1.17      \\ \hline
\end{tabular}
\end{table}

The results of the b-posit encoder follow a trend similar to that of the b-posit decoder. At 16-bit precision, the b-posit decoder draws half the power and is almost twice as fast as the posit decoder. However, it performs slightly worse than the float encoder in all three metrics.

At 32-bit precision, the b-posit encoder shows 68\% lower power usage, 46\% less area, and 44\% shorter delay compared to the posit encoder. Compared to the float encoder, it has 43\% more peak power overhead, is 8\% smaller, and has a similar delay.

At 64-bit precision, the b-posit encoder is more than twice as fast and draws one-fourth the power than the posit encoder. The delay and power of the b-posit encoder and the float encoder are almost the same at 64-bit precision. However, \textbf{the b-posit encoder is almost 32\% smaller in size}. Again, bar graphs help visualize the comparison (Fig. 15).

\begin{figure}[ht]
    \centering
    \includegraphics[width=0.5\linewidth]{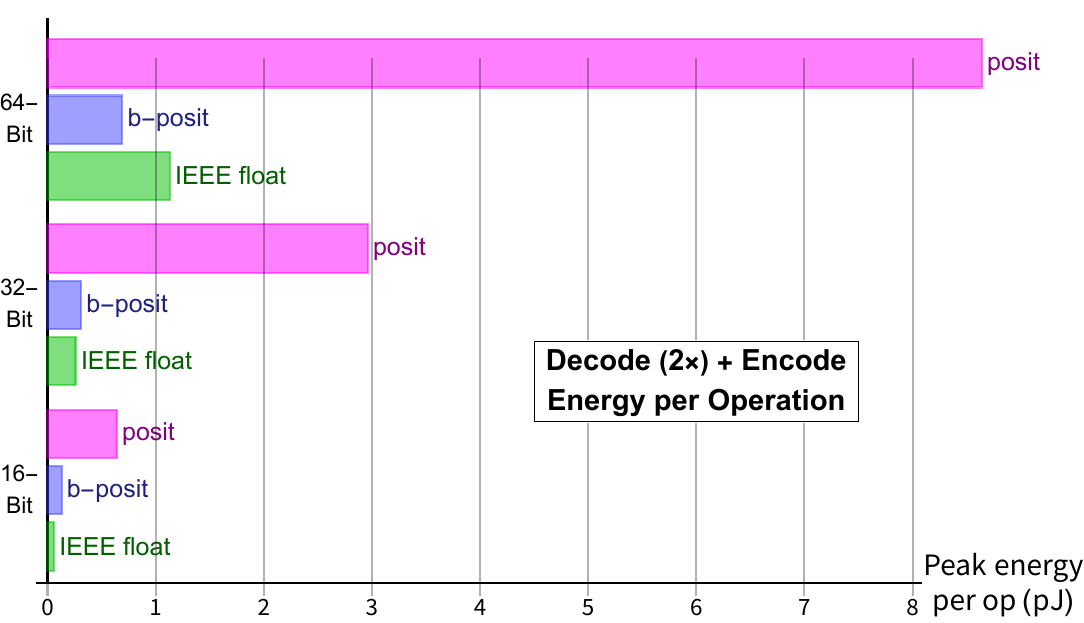}
    \caption{Worst-Case Energy Use}
    \label{fig:energycost}
\end{figure}

Note that all the data are worst-case. The power usage and delays are data-dependent for posits and b-posits, with longer regimes creating longer delays. We can compare the worst case \emph{energy use} in joules for a two-operand operation such as addition as the product of time (delay) and peak power. Assume that the two decodes are done in parallel, so the worst-case energy used is $(\mathrm{decode\_time}+\mathrm{encode\_delay})\times (2\times\mathrm{decode\_power}+\mathrm{encode\_power})$. Fig.~16 compares the three formats and three precisions for the worst-case energy consumption in pJ. For 32-bit precision, the b-posits are tied to IEEE floats. For 64-bit precision, \textbf{b-posits use 40\% less energy than IEEE floats}, and much less than the standard posit format.

\section{Conclusion}
The 2017 paper that introduced posit arithmetic to the world was titled ``Beating Floating Point at its Own Game: Posit Arithmetic''. But while $n$-bit posits beat $n$-bit floats for accuracy, dynamic range, and elegant simplicity in general, it was only speculation that $n$-bit posits would beat $n$-bit floats in power and energy usage, silicon area requirements, and delay (latency). In the years of optimizing posit hardware that followed, it looked like the idea would always have to use the fallback argument that an $n$-bit posit can often do the work of an $2n$-bit float and thereby save in every hardware cost measure.

\textbf{That has now changed}. This paper could have been titled, ``Beating Floating Point at its Own Game with the Same Precision: B-Posit Arithmetic'' because that now appears to have been achieved. Just by comparing the decode-encode overhead of float64 and b-posit64, b-posit uses less power and has less delay (which together imply less energy per operation) and requires substantially less chip area. B-posits also preserve the hard-won theorems of numerical analysis that have been amassed over many decades, since they have a \textit{minimum number of significant bits} as is required to prove error bounds on calculations. The original regime concept was sound as a way of creating tapered relative accuracy; it simply needed a tweak, that of bounding its length, for posits to achieve the goal of beating floating-point across the board.

The b-posit was developed to combine the advantages of the posit representation with the parallel decoding of floats. The b-posit maintains an effective balance between dynamic range and precision for technical workloads, while also simplifying the hardware implementation of decode–encode. The results make a case for b-posits to be used in mainstream applications with significant advantages in decode-encode compared to both standard posits and standard floats, especially at higher precisions. As examples of use on various workloads are amassed in the years to come, there may well be a case for including b-posits in the community-maintained Posit Standard. 

%
%
%
%

\end{document}